# The honeycomb and hyperhoneycomb polymorphs of IrI$_3$

Danrui Ni[a], Kasey P. Devlin[a], Guangming Cheng[b], Xin Gui[a], Weiwei Xie[c], Nan Yao[b], and Robert J. Cava*[a]

[a]Department of Chemistry, Princeton University, Princeton, NJ 08544, United States
[b]Princeton Institute for the Science and Technology of Materials, Princeton University, Princeton, NJ 08544, United States
[c]Department of Chemistry and Chemical Biology, Rutgers University, Piscataway, NJ 08854, USA

**Abstract**

The synthesis of IrI$_3$ at high pressure in its layered honeycomb polymorph is reported. Its crystal structure is refined by single crystal X-ray diffraction. Faults in the honeycomb layer stacking are observed by single crystal diffraction, synchrotron powder diffraction, and transmission electron microscopy. A previously unreported "hyperhoneycomb" polymorph of IrI$_3$ ($\beta$-IrI$_3$), is also described - Its structure in space group F*ddd* is determined by single crystal XRD. Both materials are highly-resistive diamagnetic semiconductors, consistent with a low spin $d^6$ configuration for Ir(III). The two- and three-dimensional Ir arrays in these polymorphs of IrI$_3$ are analogous to those found in the $\alpha$- and $\beta$- polymorphs of Li$_2$IrO$_3$, although the Ir electron configurations are different.

## Introduction

Heavy transition metal compounds, especially heavy metal trihalides with honeycomb layered structures, have attracted much attention in recent years.[1–4] They have been synthesized and studied as promising quantum materials, including quantum spin liquids (e.g. Kitaev-type), topological insulators, and superconductors [5–9]. In this chemical family, iridium trihalides, 5$d$ metal compounds with significant spin orbit coupling, are of particular interest. They are often known to crystallize in layers with a honeycomb lattice. Compared to IrCl$_3$ and IrBr$_3$, whose structures have been well-studied by single crystal X-ray diffraction (SCXRD)[10,11], the crystal structure of the iodine variant, IrI$_3$, is not well known. The limited studies on IrI$_3$ only report its possible space group (*C2/m*, which is the same as is seen for its sister compound IrBr$_3$), and the lattice parameters of the unit cell with relatively low precision[11,12]. One of the possible explanations for this is that, compared to other iridium trihalides, IrI$_3$ is relatively unstable and tends to decompose and release iodine when prepared using traditional vapor transport methods[11], and thus it is difficult to obtain single crystals of the compound for further characterization.

Here we report the successful preparation of IrI$_3$ by high pressure synthesis. The material once synthesized does not decompose at ambient pressure and temperature, enabling a more detailed exploration of its structural and physical properties. The honeycomb layered structure of *C*2/*m* IrI$_3$ (*α*-IrI$_3$), with its stacking faults, is determined and refined by SCXRD, synchrotron diffraction, and high-resolution scanning transmission electron microscopy (HR-STEM) studies; the basic magnetic, charge transport, and optical absorption properties of the compound are also characterized, revealing it to be a diamagnetic semiconductor. By changing the preparation conditions, a second polymorph of IrI$_3$, crystallizing in *Fddd* symmetry (*β*-IrI$_3$) is discovered. This has not been reported previously in the Ir-I system, but the same polymorphic behavior has been observed for IrCl$_3$[13] and Li$_2$IrO$_3$[14,15].

**Results and Discussion**

Honeycomb layered *α*-IrI$_3$, with its crystal structure in the *C*2/*m* space group ($a$ = 6.80 Å, $b$ = 11.79 Å, $c$ = 6.86 Å, $β$ = 109.5°, $Z$ = 4) is presented in Figure 1 and Table 1-2. The structure is refined with single crystal XRD from a crystal picked from the polycrystalline high-pressure product. The consistency with the bulk sample is confirmed by a Le Bail fitting of the powder XRD pattern (Figure 2A), with a very small amount of iridium metal impurity.

The results show that the IrI$_6$ coordination octahedron is close to ideal but slightly distorted, with an Ir-I distance ranging from 2.72 to 2.74 Å and an I-Ir-I angle ranging from 85.8° to 93.2° as shown in Figure 1. These octahedra form the 2D honeycomb layered lattice by sharing edges. The Ir-Ir distances in the honeycomb layer are 3.91 and 3.94 Å, and the distance between the layers is 6.72 Å. This monoclinic structure can be considered as being slightly distorted from ideal rhombohedral symmetry, as it can alternatively be described as a pseudo-rhombohedral unit cell with ABC layer stacking, with a gamma angle of 90.24° instead of 90°. This 0.24° distortion results in a lower symmetry for the system, as is commonly observed in layered di and trihalides. [3]

As revealed in the high-resolution synchrotron powder diffraction data for *α*-IrI$_3$ (Figure 2B), the peak shapes are asymmetric, displaying tails on their high angle side – an indication that there are a significant number of faults in the honeycomb layer stacking. Thus, electron diffraction images are collected on *α*-IrI$_3$ samples to better characterize the potential stacking faults present. Streaking is clearly observed along $c^*$ in Figure 2C, and two orientations, <100> and <110>, coexist in the same atomic image. This confirms the existence of interplanar defects, which is also clearly revealed in Figure 2D by HR-STEM, taken along the direction perpendicular to the layer stacking. The intensive stacking faults in the system lead to an apparent partial occupancy of Ir on the normally empty interstitial sites (Wyckoff position 2*b*) in the honeycomb during the structural refinement, as the SCXRD measurements are a positional average over the whole crystal. Therefore, the interplanar structural errors result in an "average structure" that has Ir atoms partially occupying both honeycomb sites and interstitial sites - the occupancy fractions are determined to be 80% on 2*b* (Ir1) and 60% on 4*g* (Ir2) respectively; however, due to the multiplicity differences of the 2*b* and 4*g* sites, the 2*b* site, occupied in the average structure determined by SCXRD due to the stacking faults, accounts for ~40% of the Ir total, while the honeycomb 4*g* site (The Ir2 sublattice shows the geometry of the honeycomb lattice) accounts for ~60% of the Ir total. This result is presented in Table 2. The TEM characterization (electron diffraction in Figure 2C and HR-STEM images in Figure 3) also indicates a *C*2/*m* symmetry with lattice parameters $a \approx 6.6$ Å, $b \approx 11.7$ Å, $c \approx 6.7$ Å, $β \approx 109.4°$ for this material, which is consistent with what is obtained from the SCXRD data.

By slowing down the cooling procedure and increasing the annealing temperature of the high-pressure synthesis process, another polymorph, *β*-IrI$_3$, forms in the products. Different from the layer-structured *α*-IrI$_3$ which was roughly reported in the literature[11,12], this *β*-polymorph has not been reported before in the IrI$_3$ system. By SCXRD refinement, *β*-IrI$_3$ is determined to crystallize in the *Fddd* space group, with an orthorhombic unit cell of $a$ = 7.91 Å, $b$ = 11.14 Å, $c$ = 23.44 Å, and $Z$ = 2 (Figure 4, Table 1 and Table 3). The edge-sharing IrI$_6$ octahedra have slightly shorter Ir-I bonds (around 2.67 Å) compared to *α*-IrI$_3$, and a I-Ir-I bond angle range of 85.1° to 92.4°. Distinguished from the 2D *α* phase, where the IrI$_6$ octahedra form a planar honeycomb lattice with van der Waals gap between the layers, the Ir-octahedra in *β*-IrI$_3$ have a three-dimensional connectivity resulting in a "hyperhoneycomb" network[15]. The structure can be considered as consisting of incomplete hexagonal rings, connecting in three dimensions through a rotation, along the *c*-axis, or as zigzag chains in alternating directions (along the two basal plane diagonals of the *ab* plane), which stack through the *c* direction (if these zigzag chains are not alternating and align parallel with the same orientation, 2D honeycomb planes are formed)[15]. This hyperhoneycomb structure type thus serves as an important potential candidate for realizing a 3D Kitaev model. The *β*-IrI$_3$ phase is most dominant in a sample that was annealed at 900 °C, 6 GPa for 1 hour and slowly cooled down to low temperature before depressurizing. This is confirmed by Le Bail refinement in Figure 5.

Although there have been no previous reports of the structural polytypes of IrI$_3$, similar polymorphic behavior has been observed for IrCl$_3$ and Li$_2$IrO$_3$, and in a different MX$_3$ structural family for the **A**P$_3$ (**A**= Sr, Ba and Eu) Zintl-phases for example.[16–18] The phase transition process is relatively complicated in IrCl$_3$[13], but in the Li$_2$IrO$_3$ system, repetitive annealing at high temperature makes the 2D layered *α*-Li$_2$IrO$_3$ transform into the 3D hyperhoneycomb *β*-Li$_2$IrO$_3$[15], suggesting that the *β* phase may be a high-temperature phase and more thermodynamically stable for Li$_2$IrO$_3$. This is also consistent with what we observed in the IrI$_3$ system, as the *β* variant appears to be a high-temperature phase, with a transition between the *α* and *β* polymorphs near 800 °C at 6 GPa.

Li$_2$IrO$_3$ has been particularly attractive to researchers in recent years due to its heavy spin-orbit coupling, and the presence of Ir(IV) with a $d^5$ configuration in its honeycomb and hyperhoneycomb structures. It has been heavily studied for both its structure types and has been considered as a potential Kitaev quantum material[15,19–21]. By comparing the structures of the polymorphs of IrI$_3$ and Li$_2$IrO$_3$, it is revealed that these two compounds have the same lattice arrangement of iridium atoms (Figure 6). Even though the chemical valences, and thus the spins, are different, the iridium is in the same Wyckoff site in the 2D honeycomb and 3D hyperhoneycomb networks in both materials. Iodine and oxygen atoms also occupy the same positions (4*i* and 8*j* in *C*2/*m*, or 16*e* and 32*h* in *Fddd*) and octahedrally coordinate the Ir. The Li atoms in Li$_2$IrO$_3$, which of course are not present in IrI$_3$, can be considered as occupying both the interstitial sites in the honeycombs, and the space between the honeycombs (for example the gap between the two honeycomb layers in the 2D structure). Our finding the same polymorphism in Li$_2$IrO$_3$ and IrX$_3$ (X = Cl and I) enlarges the platform for designing, fabricating, and characterizing iridium-containing compounds with crystal structures that are favored for quantum materials, especially for achieving a 3D Kitaev model material, and thus will be of future research interest.

Temperature- and field-dependent magnetic susceptibility measurements were conducted on polycrystalline IrI$_3$ samples. Both the *α*- and *β*- phases show diamagnetic behavior, with small paramagnetic upturns below 5 K (Figure 7), which may come from a minor fraction of paramagnetic impurities (e.g. about 3% of a spin = 1 impurity). This paramagnetic upturn results in the small deviation of the *M* vs. *H* curve of layered *α*-IrI$_3$ from diamagnetic behavior between -1 T and 1 T at 2 K: the diamagnetic nature of the bulk sample then dominates at stronger magnetic fields (Figure 7 inset). At higher temperatures such as 250 K, the field-dependent magnetization of *α*-IrI$_3$ exhibits straight line behavior, confirming the bulk diamagnetism. Resistivity (*ρ*) measurements were carried out on a dense polycrystalline piece of *α*-IrI$_3$ sample, and an as-made dense piece containing around 80% *β*-IrI$_3$. Both samples show semiconducting behavior, as their resistivities increase with decreasing temperature from 300 to 270 K (Figure 8A main panel). For both polymorphs the resistance of the measured samples exceeds the measurement upper limit at lower temperatures. Thus it is indicated that both polymorphs of IrI$_3$ are highly resistive semiconductors. Log(*ρ*) for both samples is plotted versus $T^{-1}$ in the inset of Figure 8A. Their transport activation energies (*E*$_a$) near ambient temperature can be calculated using the relationship:

$$\rho = \rho_0 e^{-\frac{E_a}{k_B T}} \tag{1}$$

in which *ρ*$_0$ is the pre-exponential term and *k*$_B$ is Boltzmann's constant. The obtained *E*$_a$ is 0.48 eV for honeycomb *α*-IrI$_3$ and 0.24 eV for the hyperhoneycomb *β* polymorph. These numbers are much smaller than the measured optical band gaps (which will be presented later), and thus suggest that there are electronic states in the band gaps of both IrI$_3$ compounds. Their diamagnetic behavior and high resistivities are consistent with the expectation of a low-spin 5$d^6$ electron configuration in Ir(III)I$_3$, and their semiconducting natures are also confirmed by DFT calculations, which will be discussed below.

To better characterize the energy gaps between filled and empty states, diffuse reflectance measurements were carried out on polycrystalline α- and β-IrI$_3$-dominated samples. It is interesting that the two polymorphs have similar absorption properties (Figure 8B), resulting in a close absorption range and comparable band gap values. The band gap values are calculated using the equation:[22,23]

$$Ah\nu = K\left(h\nu - E_g\right)^n \tag{2}$$

where *A* is the pseudo absorbance, and *K* is a constant. For direct transitions, *n* = 2; and for indirect transitions, *n* = 0.5. According to the calculated electronic band structures (discussed in a later section), the indirect transition calculation is adopted for both α- and β-IrI$_3$ (as shown in the Tauc plot in Figure 8B inset). The band gap values are 1.49 eV for α-IrI$_3$, and 1.51 eV for β variant respectively, which are consistent with the dark color of the powders. If the absorbance is due to direct transitions instead, then the band gaps not significantly different from these, being around 1.49 eV for both polymorphs.

The electronic band structures for both α- and β-IrI$_3$, calculated by DFT both without and with spin-orbit coupling (SOC) included, are presented in Figure S1 and Figure 9 respectively. The corresponding density of states figures are included as well. The hybridization of the *d*-orbitals from Ir with the *s* and *p* orbitals from I dominates the electronic states near the Fermi level (E$_F$) in both cases. When taking SOC into consideration, there is no large difference observed in the band structure of the compounds near E$_F$, with only a moderate variation in the valence bands at some points in the Brillouin Zone. Both IrI$_3$ compounds are calculated to be indirect gap semiconductors with a band gap at around 1 eV, while the band gap of β-IrI$_3$ is slightly larger than that of α-IrI$_3$.

**Conclusion**

IrI$_3$ has been successfully synthesized by a solid-state high-pressure method. The crystal structure of a previously reported monoclinic honeycomb layer phase (α) is refined by SCXRD on a sample made at 800 °C and 6 GPa. This phase has a significant number of stacking faults between its honeycomb layers that are characterized by synchrotron powder diffraction, TEM electron diffraction and STEM imaging. This highly defective polymorph suggests that the preference for three-layer stacking over two-layer stacking is relatively weak. By changing to a higher synthesis temperature, another polymorph of IrI$_3$ (the β phase), which has not been previously reported is found, this one with an orthorhombic 3D hyperhoneycomb structure. Physical and optical absorption properties are characterized on both of the IrI$_3$ polymorphs, revealing their diamagnetic behavior, high resistivities and indirect band gaps, consistent with a low spin *d$^6$* configuration of the Ir(III). The polymorphic behavior, which is newly discovered in

IrI$_3$ but similar to that seen in the well-known Li$_2$IrO$_3$ system, provides a promising platform for the fabrication and study of more potential candidates for 2D and 3D Kitaev-model magnetism, and therefore may be of future interest in quantum materials research.

**Acknowledgements**

This research was funded by the Gordon and Betty Moore foundation, EPiQS initiative, grant GBMF-9066. The single crystal diffraction work at Rutgers University was supported by U.S. DOE-BES under Contract DE-SC0022156. Use of the Advanced Photon Source for the high-resolution synchrotron experiment at Argonne National Laboratory was supported by the US Department of Energy, Office of Science, Office of Basic Energy Sciences, under Contract No. DE-AC02-06CH11357. The authors also acknowledge use of Princeton University's Imaging and Analysis Center, which is partially supported by the Princeton Center for Complex Materials (PCCM), a National Science Foundation (NSF)-MRSEC program (DMR-2011750).

**Experimental**

**Material Synthesis:** The IrI$_3$ compounds were prepared by a solid-state high pressure synthesis method. Stoichiometric elemental iridium (Alfa Aesar 99.9%) and iodine (Alfa Aesar 99.99+%, resublimed for purification) were well mixed and loaded in a boron nitride crucible, which was then inserted into a pyrophyllite cube assembly. The system was pressed to 6 GPa in a cubic multi-anvil system (Rockland Research Corporation) and heated to 700 - 900 °C at 50 °C/min where it was held for 1 hour. (The temperature was determined by an internal thermocouple.) Monoclinic layered honeycomb structure IrI$_3$ (α-IrI$_3$) was obtained by quench cooling from 700 or 800 °C. Cooling at 50 °C/min from 800 °C, or either slowly cooling or quenching from 900 °C, lead to the formation of the orthorhombic IrI$_3$ phase (β-IrI$_3$) in high proportion in the products. These results suggest that the transition between the α and β variants of IrI$_3$ is near 800 °C at 6 GPa. The resulting products are polycrystalline, but small crystals could be found for Single Crystal X-ray Diffraction (SCXRD) measurements.

**Characterization** The SCXRD data were collected on a Bruker D8 Quest Eco using graphite monochromated Mo Kα radiation (λ = 0.71073 Å). The frames were integrated using the SAINT program within the APEX III version 2017.3-0. The structures were determined using direct methods and difference Fourier synthesis (SHELXTL version 6.14).[24] The orthorhombic *Fddd* (No. 70) β-IrI$_3$ structure determination was straightforward with XPREP immediately suggesting the *Fddd* (No. 70) space group. The monoclinic *C*2/*m* (No. 12) α-IrI$_3$ structure was originally suggested as having a Trigonal *R*-3*m* (No. 166) unit cell by XPREP; however, the solution to the *R*-3*m* (No. 166) cell did not satisfactorily match the powder X-ray diffraction data or the transmission electron microscopy results. The program Cell Now was then used to identify a Monoclinic *C*2/*m* (No. 12) unit cell, which fit the powder diffraction data, and the resulting structural solution is presented.

Laboratory powder X-ray diffraction (PXRD) patterns were collected using a Bruker D8 Advance Eco with Cu Kα radiation (λ= 1.5418 Å). The Le Bail fitting of the acquired PXRD patterns was conducted via the TOPAS software. The high-resolution synchrotron powder diffraction data were acquired at Beamline 11-BM at the Advanced Photon Source at Argonne National Laboratory at a wavelength of 0.4582 Å.

For scanning transmission electron microscopy characterization, thin samples were prepared by focused ion beam cutting. All samples were polished using a 2-kV gallium ion beam to minimize the surface damage caused by the high-energy focused ion beam. Transmission electron microscopy (TEM) imaging and atomic-resolution high-angle annular dark-field scanning transmission electron microscopy imaging were performed on a Titan Cubed Themis 300 double Cs-corrected scanning/transmission electron microscope equipped with an extreme field emission gun source operated at 300 kV with a super-X energy-dispersive spectrometry system.

Temperature-dependent Magnetization (M) data were collected on a Quantum Design PPMS (Dynacool) under an applied field of 0.1 Tesla (T), using a vibrating sample magnetometer (VSM) option. The magnetic susceptibility was defined as M/H. Resistivity measurements were carried out on QD PPMS Dynacool, with platinum wires attached to samples using DuPont 4922N silver paint.

The diffuse reflectance data were collected at ambient temperature on a Cary 5000i UV-VIS-NIR spectrometer equipped with an internal DRA-2500 integrating sphere, with the reflectance data transferred to pseudo absorbance using Kubelka-Munk theory.

**Electronic Structure Calculations** The electronic structure and electronic density of states (DOS) for both forms of $IrI_3$ were calculated using the WIEN2k program, which employs the full-potential linearized augmented plane wave method (FP-LAPW) with local orbitals implemented.[25,26] The electron exchange-correlation potential used to treat the electron correlation was the generalized gradient approximation.[27] The conjugate gradient algorithm was applied. Reciprocal space integrations were completed over a 6×2×6 and 5×5×5 Monkhorst-Pack $k$-point mesh for space groups $C2/m$ and $Fddd$, respectively.[28] Spin-orbit coupling (SOC) effects were applied for both Ir and I atoms. The structural lattice parameters obtained from experiment were used in all calculations. With these settings, the calculated total energy converged to less than 0.1 meV per atom.

**Table 1**. Crystal Data for the Honeycomb (monoclinic) and Hyperhoneycomb (orthorhombic) polymorphs of $IrI_3$.

| Compound | α-$IrI_3$ | β-$IrI_3$ |
|---|---|---|
| Empirical Formula | $IrI_3$ | $Ir_8I_{24}$ |
| Temperature | 298(2) K | 299(2) K |
| Crystal System | Monoclinic | Orthorhombic |
| Space Group | $C2/m$ (12) | $Fddd$ (70) |
| Unit Cell Dimensions | $a$ = 6.8022(19) Å<br>$b$ = 11.791(3) Å<br>$c$ = 6.864(2) Å<br>$β$ = 109.528(10)° | $a$ = 7.9085(3) Å<br>$b$ = 11.1351(4) Å<br>$c$ = 23.4441(8) Å |
| Volume | 518.9(2) Å$^3$ | 2064.53(13) Å$^3$ |
| Z | 4 | 2 |
| Density (calculated) | 7.334 g/cm$^3$ | 7.373 g/cm$^3$ |
| Absorption Coefficient | 43.393 mm$^{-1}$ | 43.624 mm$^{-1}$ |
| F(000) | 944 | 3776 |
| Crystal Size | 0.038 x 0.026 x 0.020 mm$^3$ | 0.048 x 0.020 x 0.015 mm$^3$ |
| Theta range for data collection | 3.149° to 27.485° | 3.277° to 26.388° |
| Index Ranges | -8≤h≤8, -15≤k≤13, -8≤l≤8 | -9≤h≤9, -13≤k≤13, -29≤l≤29 |
| Reflections Collected | 3076 | 26037 |
| Independent Reflections | 631 [R(int) = 0.0804] | 536 [R(int) = 0.0475] |
| Completeness to Theta = 25.242° | 100.00% | 100.00% |
| Max. and Min. Transmission | 0.4249 and 0.2909 | 0.2936 and 0.1979 |
| Data / Restraints / Parameters | 631 / 0 / 25 | 536 / 0 / 21 |
| Goodness-of-fit on F$^2$ | 1.013 | 1.308 |
| Final R Indices [I > σ(I)] | R1 = 0.0518, wR2 = 0.1021 | R1 = 0.0163, wR2 = 0.0312 |
| R Indices (all data) | R1 = 0.1339, wR2 = 0.1270 | R1 = 0.0200, wR2 = 0.0321 |
| Extinction Coefficient | N/A | 0.000080(3) |
| Largest Diff. Peak and Hole | 2.710 and -2.443 e.A$^{-3}$ | 1.231 and -1.761 e.A$^{-3}$ |

**Table 2.** Atomic coordinates (x $10^4$), equivalent isotropic displacement parameters ($Å^2$ x $10^3$), Wycoff positions, and occupancies for monoclinic α-IrI$_3$. U(eq) is defined as one third of the trace of the orthogonalized $U^{ij}$ tensor. Ir1 (2b) appears to occupy the empty sites in the honeycomb and Ir2 (4g) forms the honeycomb lattice. Due to the multiplicity differences, the 2b site accounts for ~40% of the Ir total while the 4g site accounts for ~60% of the Ir total.

| Atoms | Wycoff Sites | Occupancy | x | y | z | U(eq) |
|---|---|---|---|---|---|---|
| Ir1 | 2b | 0.816 | 5000 | 0 | 0 | 12(1) |
| Ir2 | 4g | 0.595 | 0 | 1657(3) | 10000 | 20(1) |
| I3 | 4i | 1 | 2497(4) | 0 | 2204(4) | 23(1) |
| I4 | 8j | 1 | 2581(3) | 1638(2) | 7684(3) | 27(1) |

**Table 3.** Atomic coordinates (x $10^4$), equivalent isotropic displacement parameters ($Å^2$ x $10^3$), and Wycoff positions for Orthorhombic β-IrI$_3$. U(eq) is defined as one third of the trace of the orthogonalized $U^{ij}$ tensor.

| Atoms | Wycoff Sites | x | y | z | U(eq) |
|---|---|---|---|---|---|
| Ir1 | 16g | 3750 | 3750 | 2915(1) | 10(1) |
| I2 | 16e | 6042(1) | 3750 | 3750 | 15(1) |
| I3 | 32h | 3822(1) | 1357(1) | 2871(1) | 15(1) |

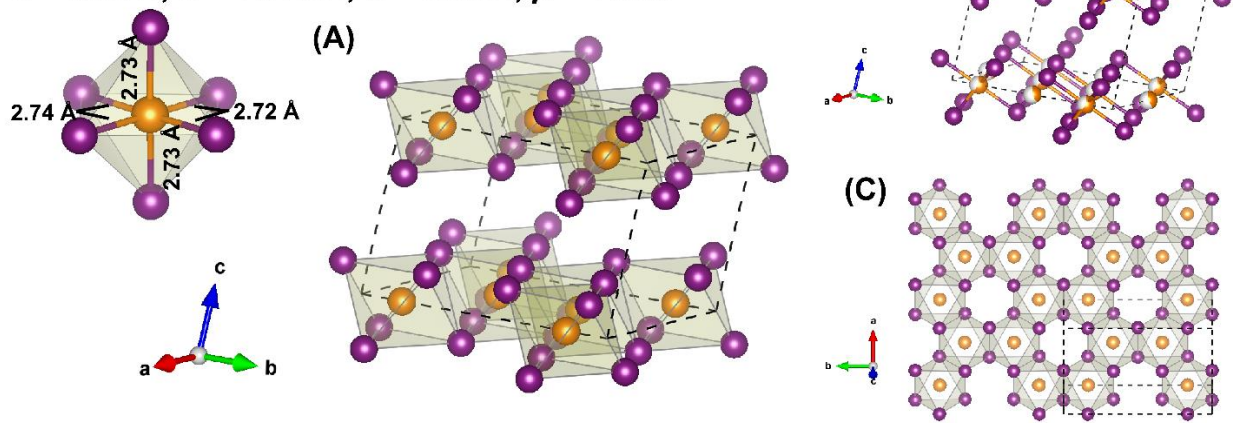

**Figure 1. The crystal structure of the 2D honeycomb layer monoclinic polymorph α-IrI₃** with an IrI$_6$ octahedron shown in the upper left corner. (A) The *C*2/*m* unit cell with (B) the average structure refined from single crystal XRD, and (C) a view of the honeycomb plane. Orange spheres represent iridium and purple spheres represent iodine.

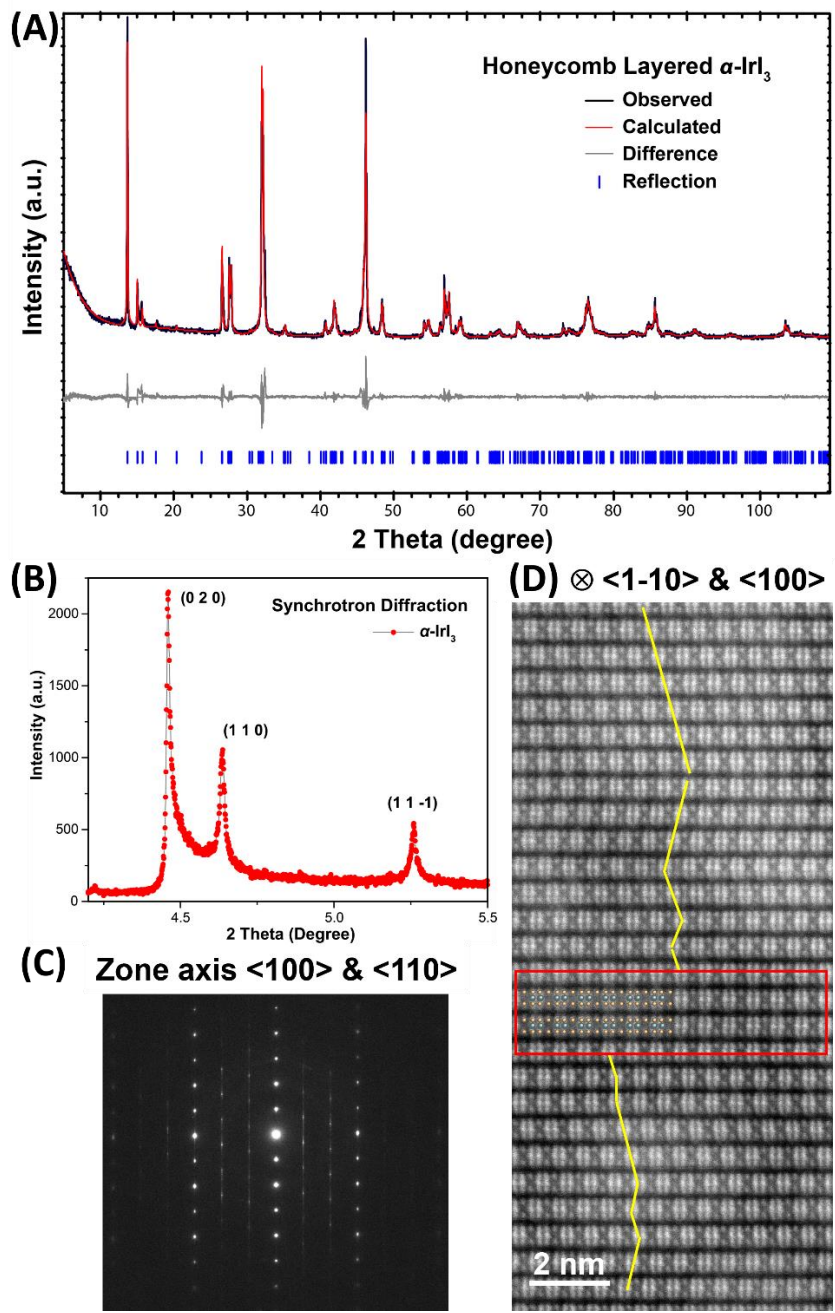

**Figure 2. The crystal structure of the layered polymorph of IrI$_3$** (A) Le Bail fit of the laboratory PXRD pattern of 2D honeycomb layered α-IrI$_3$, Rwp% = 13.8, GOF = 2.02; (B) Partial synchrotron powder diffraction pattern of α-IrI$_3$, showing the asymmetric peak shapes that are characteristic of stacking faults; (C) Electron diffraction image of α-IrI$_3$, collected with the beam parallel to the <110> direct lattice direction. The streaking pattern in the vertical direction comes from the stacking faults; (D) An atomic resolution STEM image showing the stacking faults between honeycomb layers from zone axis <1-10> (as indicated by the zigzagging yellow line), and the stacking from <100> is marked with red rectangle.

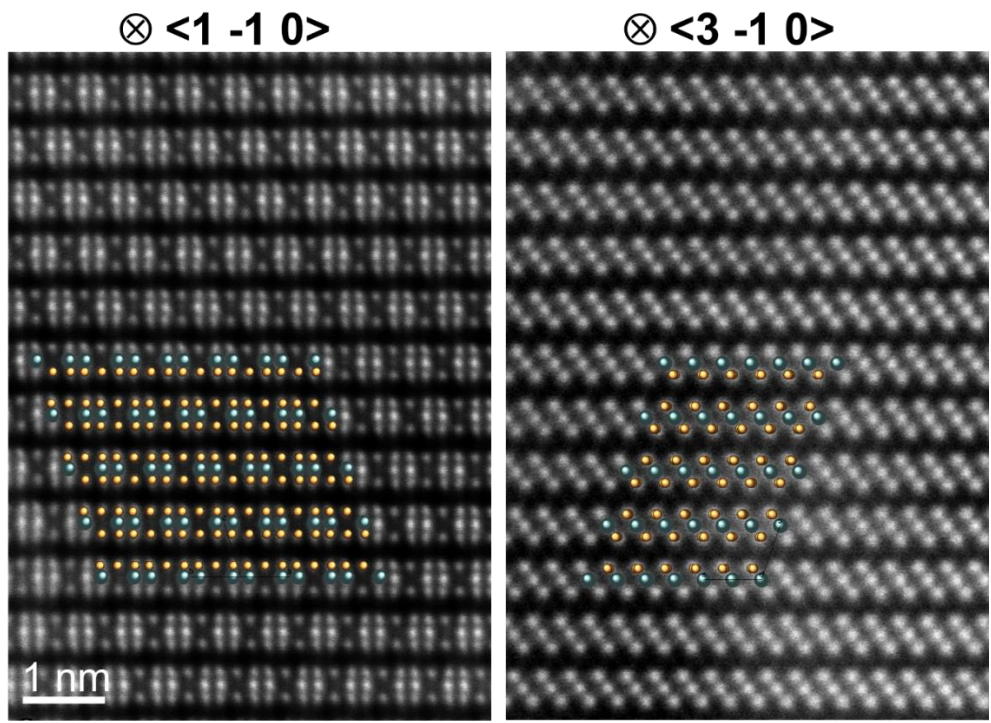

**Figure 3. Atomic-resolution STEM images of α-IrI$_3$** showing the atomic staking. The two images are taken with the beam parallel to the <1-10> and <3-10> directions, respectively. Representative atomic positions are marked with orange (iodine) or blue (iridium) colored dots on the STEM images.

# Hyperhoneycomb β-IrI₃
*Fddd* (#70), *a* = 7.91 Å, *b* = 11.14 Å, *c* = 23.44 Å

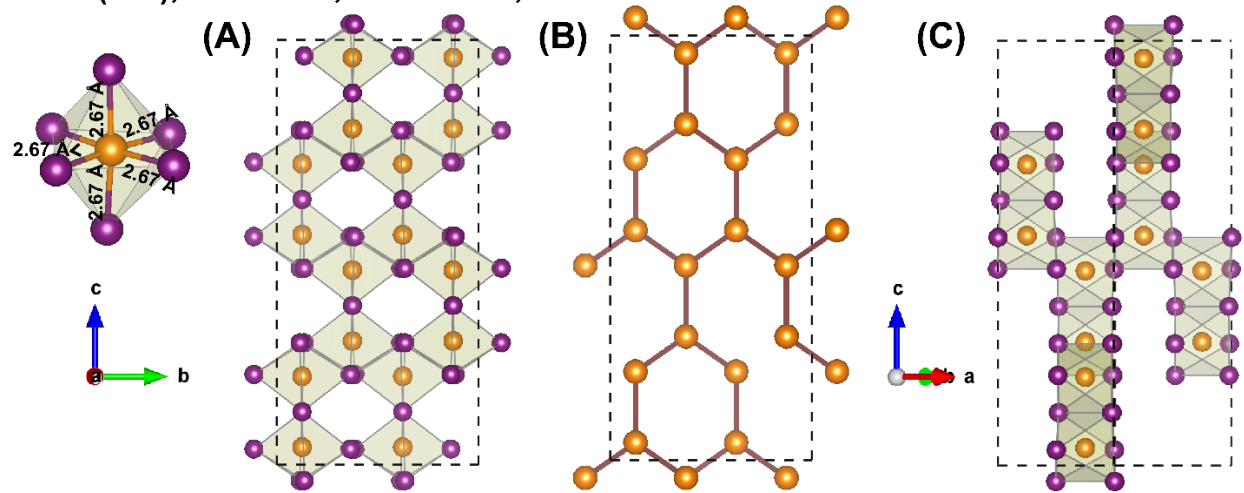

**Figure 4**. **The crystal structure of the 3D hyperhoneycomb polymorph of *β*-IrI₃** with its IrI₆ octahedron shown on the upper left corner. (A) The *Fddd* unit cell is shown, together with (B) the view with only Ir atoms presented in order to exhibit the hyperhoneycomb arrangement. The three-dimensional connection of the hyperhoneycombs is revealed in (C) along a direction perpendicular to the *c*-axis. Orange colored spheres represent iridium and purple ones represent iodine.

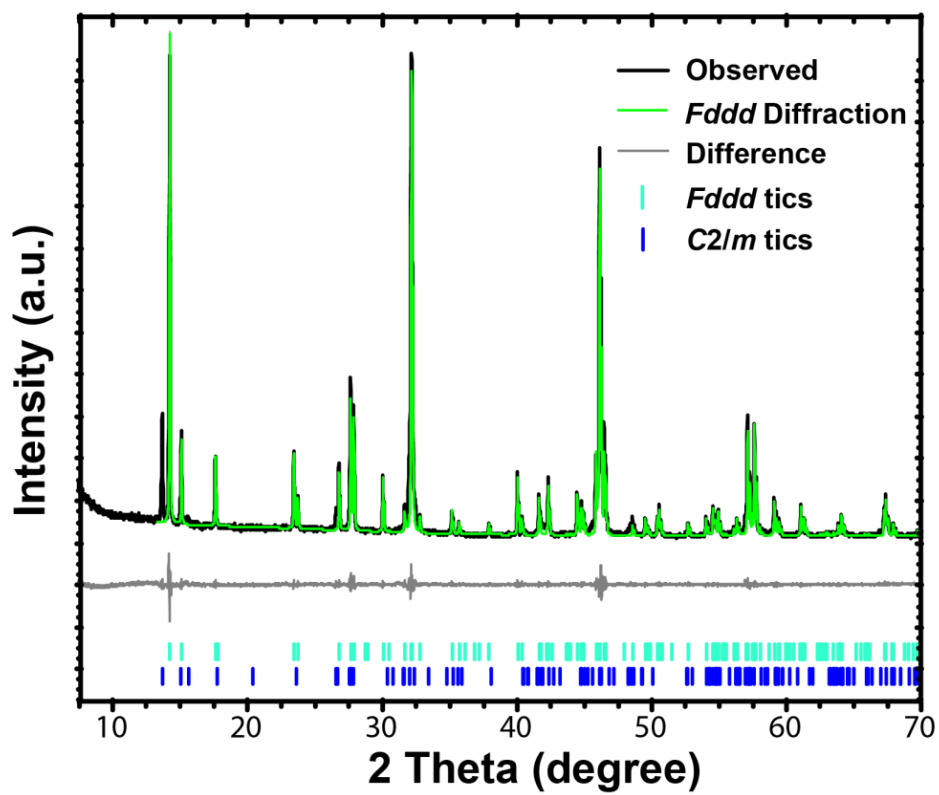

**Figure 5. Powder diffraction of a sample dominated by the hyperhoneycomb polymorph β-IrI$_3$**
Le Bail fit of the laboratory PXRD pattern, collected on a 900 °C annealed, slowly cooled sample of IrI$_3$. Rwp = 8.98, GOF = 1.31. The *Fddd* polymorph β-IrI$_3$ (green colored curve, cyan colored tics) dominates in this sample, with a small amount of α-IrI$_3$ coexisting (blue tics).

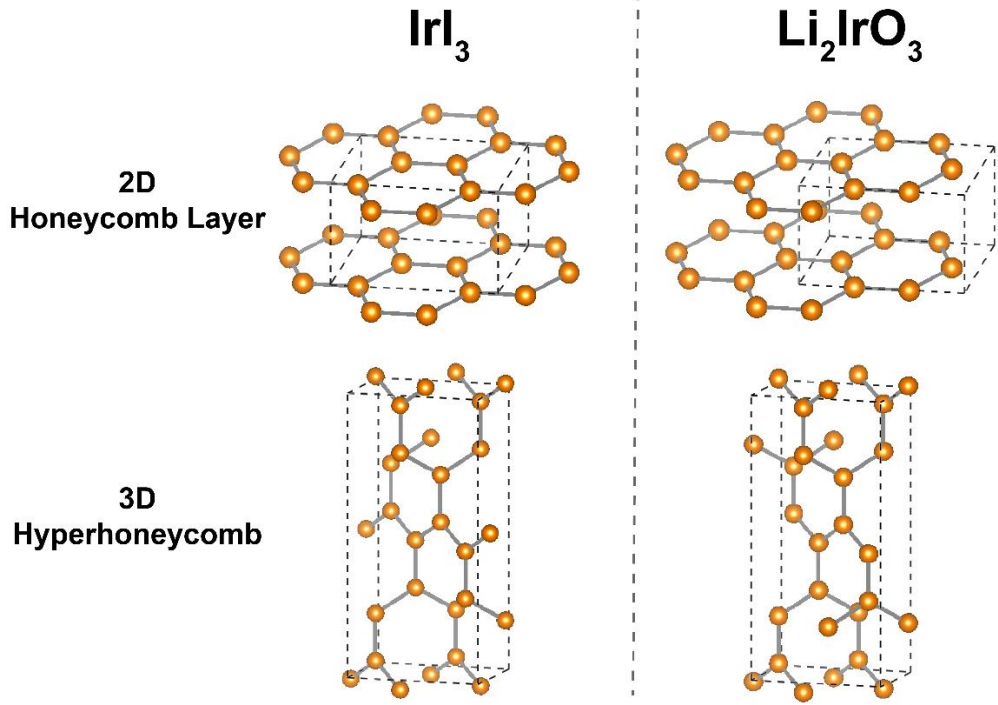

**Figure 6. Structural comparison between the IrI₃ (left) and Li₂IrO₃ (right) polymorphs**, with the Ir atomic arrangement emphasized. The 2D layered honeycomb structure (*C*2/*m*) is on top and the 3D hyperhoneycomb (*Fddd*) polymorph is on the bottom. The Iodine atoms in IrI₃ and the lithium and oxygen atoms in Li₂IrO₃ are omitted.

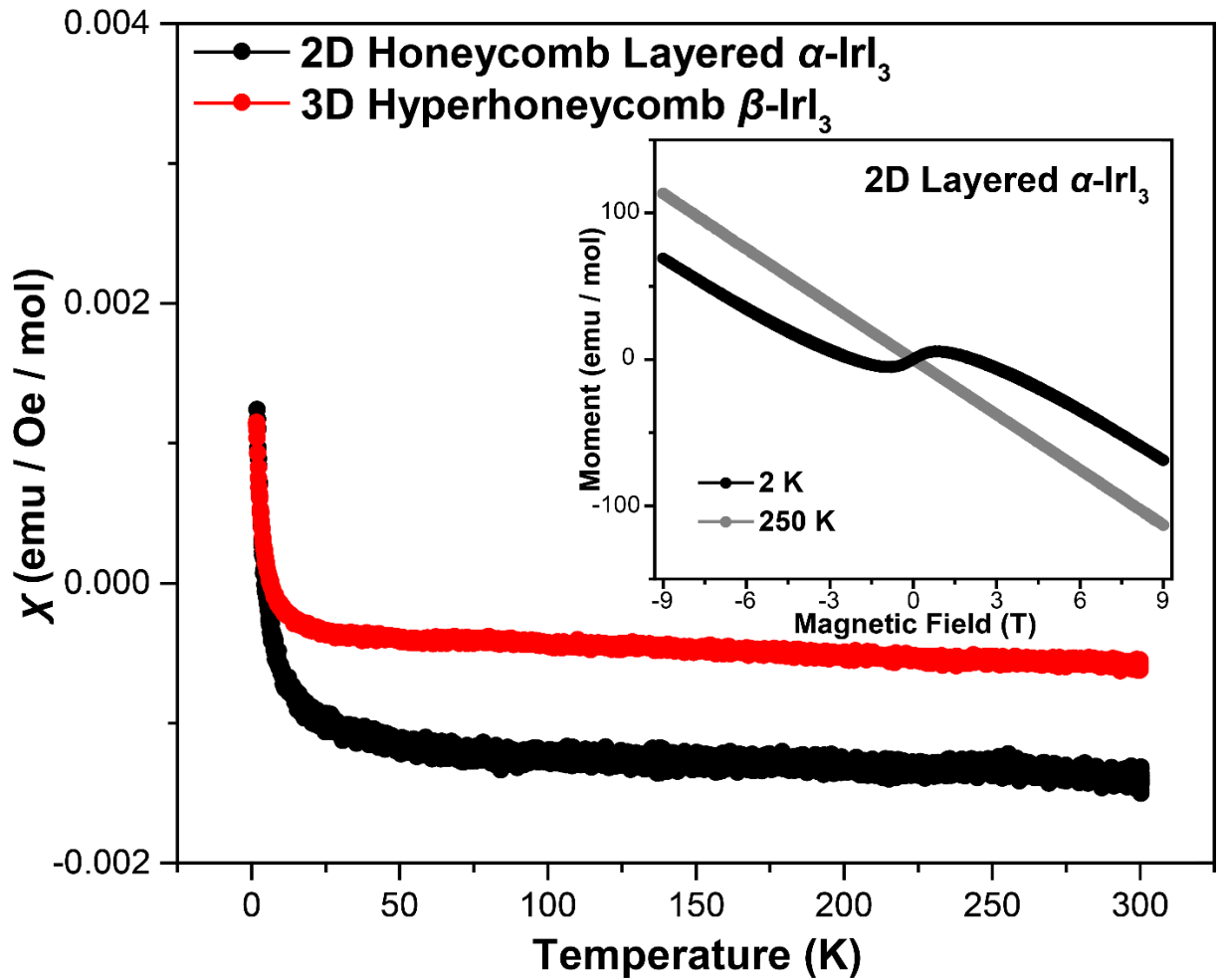

**Figure 7. Temperature-dependent magnetic susceptibilities of *α*- and *β*-IrI₃**, measured from 1.8 to 300 K. The inset presents field-dependent magnetization measurements of polycrystalline *α*-IrI₃ powder from -9 T to 9 T at two different temperatures.

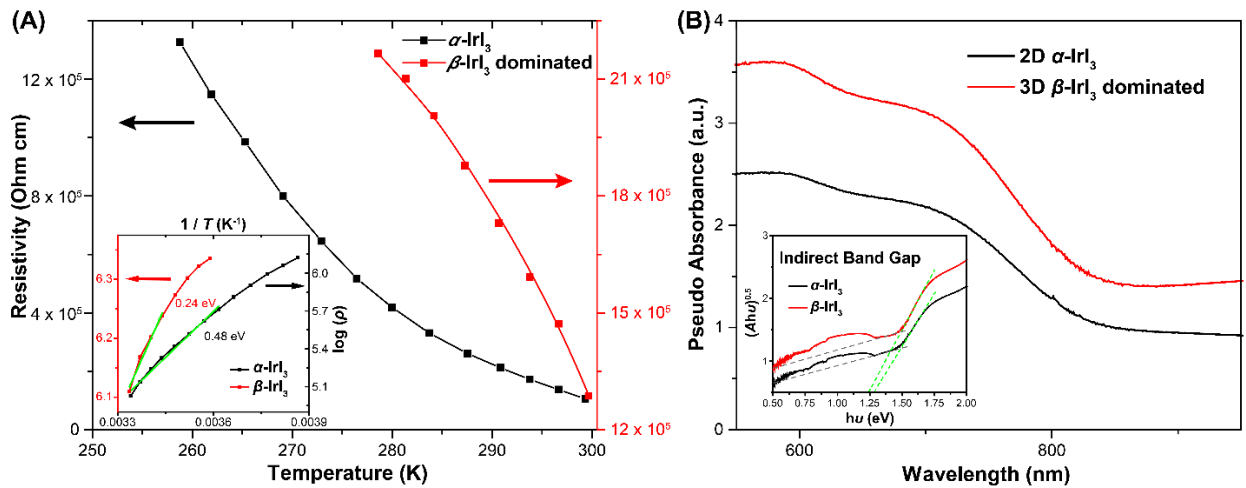

**Figure 8. Electronic transport property and optical absorption characterization of IrI$_3$.** (A) Resistivity measured from 260 to 300 K on an as-made dense piece of α-IrI$_3$ (left *y*-axis), and from 270 to 300 K on an as-made dense sample which is dominated by β-IrI$_3$ (right *y*-axis). The inset shows the log(*ρ*) of both samples plotted versus *T*$^{-1}$, with the linear fitting (green lines) on the higher temperature range of each curve to calculate the activation energy. (B) Pseudo Kubelka-Munk absorbance spectrum (main panel) and the Tauc plot for the indirect optical transition (inset) of polycrystalline IrI$_3$ samples. The optical bandgap is estimated (in the inset) by extrapolating the intersection of the green dashed line (the linear absorption region) and the gray dashed line (absorption baseline).

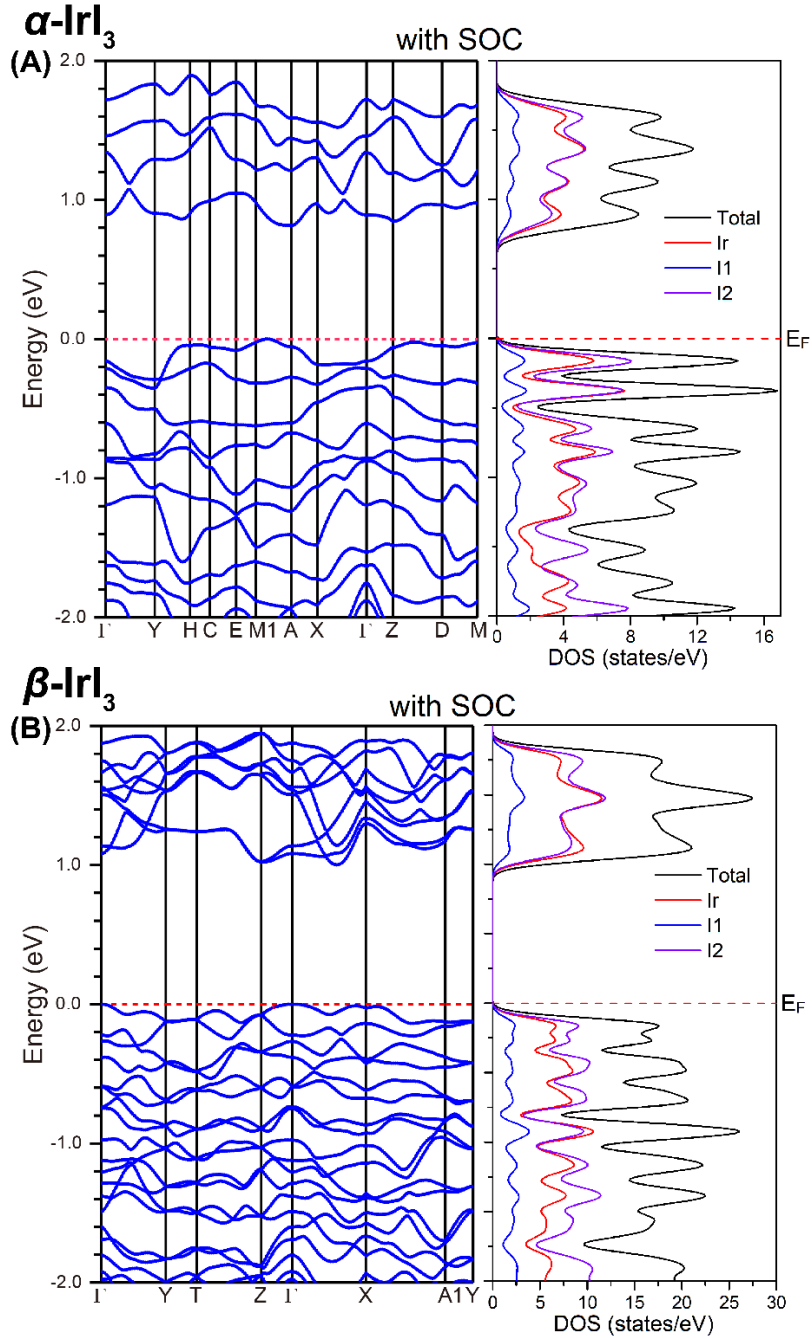

**Figure 9. DFT-calculated electronic band structures** with spin orbit coupling (SOC) included, for (A) *α*- and (B) *β*-IrI$_3$. The electronic density of states (DOS) figures are presented next to the corresponding band structures. The contributions of different atoms to the DOS are labeled in different colors. I1 refers to the iodine on the 4*i* site of *α*-IrI$_3$, and the 16*e* site of *β*-IrI$_3$; I2 refers to I on the 28*j* site in *α* and I on the 32*h* site in *β*.